\documentclass[aps,prl,preprint,tightenlines,superscriptaddress,showpacs]{revtex4}
\def\mystyle{4}

\if\mystyle4
 \def\figonescale{0.7}
 \def\figtwoscale{0.8}
 
  \def\bellelogo{\vbox to 16mm{
                 \vss\hbox{\resizebox{!}{3cm}{
                 \includegraphics{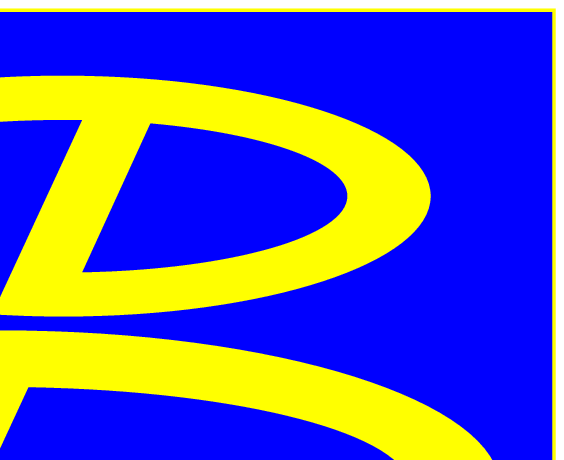}}}}\vspace{-1cm}}
  \def\preprintA{\hbox{\hfil BELLE-CONF-0401}}
  \def\preprintB{\hbox{\hfil ICHEP 11-0646}}
  \def\preprintC{}
\else
 \def\figonescale{0.4}
 \def\figtwoscale{0.2}
 
 \def\bellelogo{}
 \def\preprintA{}
 \def\preprintB{}
 \def\preprintC{}
\fi

\def\mydate{\date{August 18 2004}}

\usepackage{graphicx}

\def\KM{K^-}

\def\KS{{K^0_S}}

\def\piZ{{\pi^0}}
\def\piP{{\pi^+}}
\def\piM{{\pi^-}}

\def\rhoM{{\rho^-}}
\def\rhoZ{{\rho^0}}
\def\Kstar{{K^*}}

\def\KstarM{{K^{*-}}}
\def\KstarB{{\Kbar^{*0}}}
\def\omegaG{{\omega\gamma}}

\def\rhoMG{{\rhoM\gamma}}
\def\rhoZG{{\rhoZ\gamma}}
\def\ROG{{(\rho,\omega)\gamma}}
\def\KstarG{{\Kstar\gamma}}

\def\KstarMG{{\KstarM\gamma}}

\def\qqbar{q\overline{q}{}}

\def\Bbar{\overline{B}{}}

\def\Kbar{\overline{K}{}}

\def\epem{e^+e^-{}}

\def\btosgamma{b\to s\gamma}

\def\btodgamma{b\to d\gamma}

\def\BtoRG{B\to \rho\gamma}

\def\BtoRMG{B^-\to \rho^-\gamma}
\def\BtoRZG{B^0\to \rho^0\gamma}
\def\BtoRBG{\Bbar^0\to \rho^0\gamma}
\def\BtoROG{B\to (\rho,\omega)\gamma}
\def\BtoOG{\Bbar^0\to \omega\gamma}

\def\BtoKG{B\to K^*\gamma}

\def\BtoXsgamma{B\to X_s\gamma}

\def\BtoKBG{\Bbar^0\to \Kbar^{*0}\gamma}
\def\BtoKMG{B^-\to K^{*-}\gamma}

\def\BtoDZpi{B^-\to D^0\pi^-}

\def\cm{\mbox{~cm}}

\def\GeV{\mbox{~GeV}}
\def\GeVc{\mbox{~GeV}/c}
\def\GeVcc{\mbox{~GeV}/c^2}

\def\MeVc{\mbox{~MeV}/c}
\def\MeVcc{\mbox{~MeV}/c^2}

\def\Vtd{V_{td}}
\def\Vts{V_{ts}}

\def\Br{{\cal B}}

\def\Lpi{L_\pi}
\def\LK{L_K}
\def\LpiK{{\cal L}_{\pi/K}}

\def\Mbc{M_{\rm bc}}
\def\DeltaE{\Delta{E}}
\def\Ebeam{E^*_{\rm beam}{}}

\def\MKpi{M_{K\pi}}

\def\Egamma{E_\gamma}

\def\tauBratio{{\tau_{B^+}\over\tau_{B^0}}}

\def\piZeta{\piZ/\eta}

\def\thetaB{{\theta^*_B}}
\def\cosB{{\cos\thetaB}}

\def\calF{{\cal F}}
\def\calLs{{\cal L}_s}
\def\calLc{{\cal L}_c}
\def\calR{{\cal R}}
\def\calL{{\cal L}}

\def\Deltaz{\Delta{z}}
\def\thetahel{\theta_{\rm hel}}
\def\coshel{\cos\thetahel}

\def\Lzero{{\cal L}_0}
\def\Lmax{{\cal L}_{\rm max}}

\def\PM#1#2{\,^{+#1}_{-#2}{}}
\def\EM#1{\times10^{-#1}}

\def\etal{\textit{et al.}}
\def\Journal#1#2#3#4{{#1} {\bf #2}, #3 (#4)} 
\def\NIMA{Nucl. Instrum. Meth. A}
\def\NPB{Nucl. Phys. B}
\def\PLB{Phys. Lett. B}
\def\PRL{Phys. Rev. Lett.}
\def\PRD{Phys. Rev. D}
\def\ZPC{Z. Phys. C}

\def\EPJC{Eur. Phys. J. C}
\def\JPG{J. Phys. G}

\makeatletter
\def\mygraphic@[#1,#2]#3{\resizebox{#1}{#2}{\includegraphics{#3}}}
\def\mygraphics{\@ifnextchar[{\mygraphic@}{\mygraphic@[\textwidth,!]}}
\def\myep@[#1]#2{\resizebox{#1\textwidth}{!}{\includegraphics{#2}}}
\def\myeps{\@ifnextchar[{\myep@}{\myep@[1]}}
\makeatother

\def\EffKaon{76\mbox{--}80\%} 
\def\EffPionR{89\%}  

\def\EffPionO{94\%}  
\def\FakePionR{{\sim}10\%} 

\def\EffVtxVeryRough{85\%} 

\def\drcut{2\cm}
\def\dzcut{5\cm}
\def\ptrkcut{100\MeVc}

\def\effRPG{(5.5\pm0.4)\%}
\def\effRZG{(3.9\pm0.3)\%}
\def\effOMG{(3.9\pm0.4)\%}

\def\sROG{1.9}
\def\sRPG{2.1}
\def\sRZG{0.6}
\def\sOMG{0.2}

\def\BrROG{(0.72\PM{0.43}{0.39}\PM{0.28}{0.27})\EM6}

\def\ULROG{1.4\EM6}
\def\ULRPG{2.2\EM6}
\def\ULRZG{0.8\EM6}
\def\ULOMG{0.8\EM6}

\def\ULROGoverKG{0.035}

\def\ULVtdoVts{0.21}

\begin{document}


\bellelogo

\preprint{\vbox{
  \preprintA
  \preprintB
  \preprintC
}}

\title{Search for the \boldmath$\btodgamma$ process}

\affiliation{Aomori University, Aomori}
\affiliation{Budker Institute of Nuclear Physics, Novosibirsk}
\affiliation{Chiba University, Chiba}
\affiliation{Chonnam National University, Kwangju}
\affiliation{Chuo University, Tokyo}
\affiliation{University of Cincinnati, Cincinnati, Ohio 45221}
\affiliation{University of Frankfurt, Frankfurt}
\affiliation{Gyeongsang National University, Chinju}
\affiliation{University of Hawaii, Honolulu, Hawaii 96822}
\affiliation{High Energy Accelerator Research Organization (KEK), Tsukuba}
\affiliation{Hiroshima Institute of Technology, Hiroshima}
\affiliation{Institute of High Energy Physics, Chinese Academy of Sciences, Beijing}
\affiliation{Institute of High Energy Physics, Vienna}
\affiliation{Institute for Theoretical and Experimental Physics, Moscow}
\affiliation{J. Stefan Institute, Ljubljana}
\affiliation{Kanagawa University, Yokohama}
\affiliation{Korea University, Seoul}
\affiliation{Kyoto University, Kyoto}
\affiliation{Kyungpook National University, Taegu}
\affiliation{Swiss Federal Institute of Technology of Lausanne, EPFL, Lausanne}
\affiliation{University of Ljubljana, Ljubljana}
\affiliation{University of Maribor, Maribor}
\affiliation{University of Melbourne, Victoria}
\affiliation{Nagoya University, Nagoya}
\affiliation{Nara Women's University, Nara}
\affiliation{National Central University, Chung-li}
\affiliation{National Kaohsiung Normal University, Kaohsiung}
\affiliation{National United University, Miao Li}
\affiliation{Department of Physics, National Taiwan University, Taipei}
\affiliation{H. Niewodniczanski Institute of Nuclear Physics, Krakow}
\affiliation{Nihon Dental College, Niigata}
\affiliation{Niigata University, Niigata}
\affiliation{Osaka City University, Osaka}
\affiliation{Osaka University, Osaka}
\affiliation{Panjab University, Chandigarh}
\affiliation{Peking University, Beijing}
\affiliation{Princeton University, Princeton, New Jersey 08545}
\affiliation{RIKEN BNL Research Center, Upton, New York 11973}
\affiliation{Saga University, Saga}
\affiliation{University of Science and Technology of China, Hefei}
\affiliation{Seoul National University, Seoul}
\affiliation{Sungkyunkwan University, Suwon}
\affiliation{University of Sydney, Sydney NSW}
\affiliation{Tata Institute of Fundamental Research, Bombay}
\affiliation{Toho University, Funabashi}
\affiliation{Tohoku Gakuin University, Tagajo}
\affiliation{Tohoku University, Sendai}
\affiliation{Department of Physics, University of Tokyo, Tokyo}
\affiliation{Tokyo Institute of Technology, Tokyo}
\affiliation{Tokyo Metropolitan University, Tokyo}
\affiliation{Tokyo University of Agriculture and Technology, Tokyo}
\affiliation{Toyama National College of Maritime Technology, Toyama}
\affiliation{University of Tsukuba, Tsukuba}
\affiliation{Utkal University, Bhubaneswer}
\affiliation{Virginia Polytechnic Institute and State University, Blacksburg, Virginia 24061}
\affiliation{Yonsei University, Seoul}
  \author{K.~Abe}\affiliation{High Energy Accelerator Research Organization (KEK), Tsukuba} 
  \author{K.~Abe}\affiliation{Tohoku Gakuin University, Tagajo} 
  \author{N.~Abe}\affiliation{Tokyo Institute of Technology, Tokyo} 
  \author{I.~Adachi}\affiliation{High Energy Accelerator Research Organization (KEK), Tsukuba} 
  \author{H.~Aihara}\affiliation{Department of Physics, University of Tokyo, Tokyo} 
  \author{M.~Akatsu}\affiliation{Nagoya University, Nagoya} 
  \author{Y.~Asano}\affiliation{University of Tsukuba, Tsukuba} 
  \author{T.~Aso}\affiliation{Toyama National College of Maritime Technology, Toyama} 
  \author{V.~Aulchenko}\affiliation{Budker Institute of Nuclear Physics, Novosibirsk} 
  \author{T.~Aushev}\affiliation{Institute for Theoretical and Experimental Physics, Moscow} 
% ...
\collaboration{The Belle Collaboration}

\mydate

\begin{abstract} 

We report search results for the flavor-changing neutral current process
$\btodgamma$ with a data sample containing 274 million $B$ meson pairs
accumulated at the $\Upsilon(4S)$ resonance with the Belle detector at
KEKB.  We studied the exclusive decays $\BtoRMG$, $\BtoRBG$, and
$\BtoOG$, and find no significant signal.  We set an upper limit for a
combined branching fraction $\Br(\BtoROG)<\ULROG$ at the 90\%
confidence level, which is normalized to that of $\BtoRMG$ assuming an
isospin relation between the three modes.

\end{abstract}




\maketitle



\begin{figure}[ht]
\begin{center}
\vspace*{24pt}
\myeps[\figonescale]{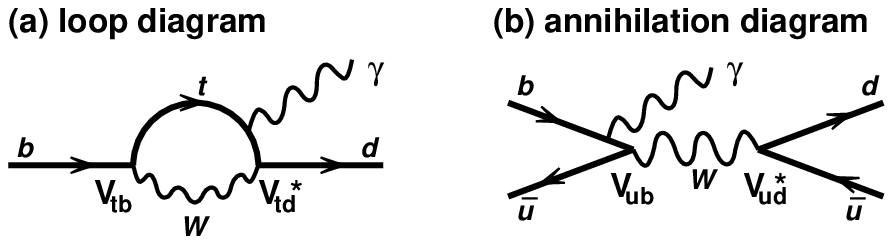}
\caption{(a) Loop diagram for $\btodgamma$ and (b) annihilation diagram
  only for $\BtoRMG$.}
\label{fig:diagram}
\end{center}
\end{figure}

The $\btodgamma$ process, shown in Fig.~\ref{fig:diagram}(a), is a
flavor changing neutral current transition that proceeds via loop
diagrams in the Standard Model (SM).  It is suppressed with respect to
$\btosgamma$ by the Cabibbo-Kobayashi-Maskawa (CKM) factor
$|\Vtd/\Vts|^2 \sim 0.04$ with a large uncertainty due to the lack of
precise knowledge of $\Vtd$.  The exclusive modes $\BtoRG$ and
$\omega\gamma$, which are presumably dominant, have not yet been
observed \cite{bib:belle-rhogam,bib:babar-rhogam}.  They are also
suppressed with respect to the corresponding exclusive decay $\BtoKG$ by
$|\Vtd/\Vts|^2$, with corrections due to form factors, $SU(3)$ breaking
effects and the additional annihilation diagrams
(Fig.~\ref{fig:diagram}(b)), giving predicted branching fractions in the
range $(0.9\mbox{--}2.7)\EM6$ in the
SM~\cite{bib:ali-parkhomenko,bib:bosch-buchalla}.  Measurement of these
exclusive branching fractions would improve the constraints on $\Vtd$ in
the context of the SM, and would provide sensitivity to physics beyond
the SM that is complementary to that from $\btosgamma$.

In this paper, we report the results of a search for the $\btodgamma$
process using a data sample of $(274\pm3)$ million $B$ meson pairs
accumulated at the $\Upsilon(4S)$ resonance.  The data are produced in
$\epem$ annihilation at the KEKB energy-asymmetric (3.5 on 8 GeV)
collider~\cite{bib:kekb} and collected with the Belle
detector~\cite{bib:belle-detector}.  The Belle detector is a
large-solid-angle spectrometer that includes a silicon vertex detector
(SVD), a central drift chamber (CDC), an array of aerogel threshold
Cherenkov counters (ACC), time-of-flight (TOF) scintillation counters,
and an electromagnetic calorimeter (ECL) comprised of CsI(Tl) crystals
located inside a superconducting solenoid coil that provides a 1.5 T
magnetic field.  An iron flux-return located outside of the coil is
instrumented to identify muons (KLM).  The dataset consists of two
subsets with different inner detector configurations: for the first 152
million $B$ meson pairs, a 2.0 cm radius beampipe and a 3-layer SVD were
used; and for the remaining 122 million $B$ meson pairs, a 1.5 cm radius
beampipe, a 4-layer SVD and a small-cell inner drift chamber were
used~\cite{bib:svd2}.


We reconstruct the following final states: $\BtoRMG$, $\BtoRBG$, and
$\BtoOG$.  (Charge conjugate modes are implied throughout this paper.)
We also reconstruct control samples of $\BtoKMG$ and $\BtoKBG$ decays.
The following decay chains are used to reconstruct the intermediate
states: $\rhoM\to\piM\piZ$, $\rhoZ\to\piP\piM$, $\omega\to\piP\piM\piZ$,
$\KstarM\to\KM\piZ$, $\KstarB\to\KM\piP$, and $\piZ\to\gamma\gamma$.

Photon candidates are reconstructed from isolated clusters in the ECL
that have no corresponding charged track, and a shower shape that is
consistent with that of a photon.  The photon with the largest
center-of-mass (CM) energy in the range $1.4\GeV<\Egamma<3.4\GeV$ and in
the barrel region of the ECL ($33^\circ<\theta_\gamma<128^\circ$ in the
laboratory frame) is selected as the primary photon candidate.  To
suppress backgrounds from $\piZ\to\gamma\gamma$ and
$\eta\to\gamma\gamma$ decays, we veto the event if the reconstructed
mass of the primary photon and any other photon of 30 (200) MeV or more
is within $\pm18$ $(32)\MeVcc$ of the $\piZ$ ($\eta$) mass.  These
correspond to $\pm3\sigma$ windows, where $\sigma$ is the mass
resolution.  This set of criteria is referred to as the $\piZeta$ veto.
For the primary photon, we sum the energy deposited in arrays of
$3\times3$ cells and $5\times5$ cells around the maximum energy ECL
cell; if their ratio is less than 0.95, the event is vetoed.

Charged pions and kaons are reconstructed as tracks in the CDC and SVD.
Each track is required to have a momentum greater than $\ptrkcut$ and
closest approach to the run-averaged interaction point within $\drcut$
in radius and $\pm\dzcut$ along the $z$-axis, aligned opposite the
positron beam.  We do not use the track to form the signal candidate if,
when combined with any other track, it forms a $\KS$ candidate with a
mass within $\pm10\MeVcc$ around the nominal $\KS$ mass and a displaced
vertex that is consistent with a $\KS$.  We determine the pion ($\Lpi$)
and kaon ($\LK$) likelihoods from the ACC response, specific ionization
($dE/dx$) in the CDC and TOF flight-time measurements for each track,
and form a likelihood ratio $\LpiK=\Lpi/(\Lpi + \LK)$ to separate pions
and kaons.  We require $\LpiK > 0.85$ for pions, which gives an
efficiency of $\EffPionR$ for pions and misidentification probability of
$\FakePionR$ for kaons.  For the $\omegaG$ mode, we relax the
requirement to $\LpiK > 0.8$, which gives an efficiency of $\EffPionO$
for pions.  (For kaons in the $\KstarG$ modes, we require $\LpiK<0.4$,
which gives an efficiency of $\EffKaon$.)  In addition, we remove pion
and kaon candidates if they are consistent with being electrons based on
ECL, $dE/dx$ and ACC information, or consistent with muons based on KLM
information.

Neutral pions are formed from two photons with invariant masses within
$\pm 10$ ($16$)$\MeVcc$ of the $\piZ$ mass, corresponding to a ${\sim}2
\sigma$ $({\sim}3\sigma)$ window for $\rhoMG$ and $\KstarMG$ ($\omegaG$)
modes.  The photon momenta are then recalculated with a $\piZ$ mass
constraint.  We require each photon energy to be greater than 30 MeV.
We also require the CM momentum of the $\piZ$ to be greater than
$0.5\GeVc$ for the $\rhoMG$ and $\KstarMG$ modes.

Invariant masses for the $\rho$ and $\omega$ candidates are required to
be within windows of $\pm150\MeVcc$ ($1\Gamma$) and $\pm30\MeVcc$
($3.5\Gamma$), respectively, around their nominal masses, where $\Gamma$
is the natural width of each resonance.

We form $B$ candidates by combining a $\rho$ or $\omega$ candidate and
the primary photon using two variables: the beam-energy constrained mass
$\Mbc = \sqrt{ (\Ebeam/c^2)^2 - |p_{B}^*/c|^{2}}$ and the energy
difference $\Delta E = E^*_{B} - \Ebeam$, where $p^*_{B}$ and $E^*_B$
are the measured CM momentum and energy, respectively, of the $B$
candidate, and $\Ebeam$ is the CM beam energy.  The magnitude of the
photon momentum is replaced by $(\Ebeam - E_{\rho/\omega}^*)/c$ when the
momentum $p^*_{B}$ is calculated.  To optimize the event selection, we
count Monte Carlo (MC) events in the region $-0.10\GeV<\DeltaE<0.08\GeV$
and $5.273\GeVcc<\Mbc<5.285\GeVcc$.  We choose the selection criteria to
maximize $N_S/\sqrt{N_B}$, where $N_S$ is the expected signal yield in
this region assuming the branching fractions to be the SM value in
Ref.~\cite{bib:ali-parkhomenko}, and $N_B$ is the expected background
yield in the same region.


The dominant background arises from continuum events
($\epem\to\qqbar(\gamma)$) where the accidental combination of a $\rho$
or $\omega$ candidate with a photon forms a $B$ candidate.  We suppress
this background using an event shape discriminator $\calF$, the $B$
candidate polar angle $\thetaB$ in the CM frame, the vertex separation
$\Deltaz$, and the output of the $B$-flavor tagging algorithm:

\begin{enumerate}
\item 
The event shape discriminator $\calF$~\cite{bib:belle-pi0pi0} is a
Fisher discriminant \cite{bib:fisher} constructed from 16 modified
Fox-Wolfram moments~\cite{bib:belle-pi0pi0,bib:fox-wolfram} and the
scalar sum of the transverse momentum of all charged tracks and photons.

\item 
True $B$ mesons follow a $1-\cos^2\thetaB$ distribution in $\cosB$,
while candidates in the continuum background are uniformly distributed.

\item 
In about $\EffVtxVeryRough$ of events for the $\rhoZG$ and $\omegaG$
modes, a fit can be successfully performed to determine the decay vertex
of the candidate $B$ meson as well as the origin of the remaining tracks
in the event.  The separation $\Delta z$ between these two vertices
along the $z$-axis discriminates between continuum events, which have a
common vertex, and signal events, whose decay vertices are displaced.

\item
The $B$-flavor tagging algorithm described in Ref.~\cite{bib:hamlet}
returns the flavor of the other $B$ meson ($q=\pm1$), and a tagging
quality $r$ ($0<r<1$) which indicates the level of confidence in the
flavor determination.  The algorithm uses the particles in the event
that are not associated with the signal $B$ candidate, and provides
additional discrimination between signal, and continuum background where
no true $B$ meson is present.
\end{enumerate}

For each of the quantities $\calF$, $\cosB$ and $\Deltaz$,
we construct one-dimensional likelihood distributions for signal and
continuum.  Signal distributions are modeled with an asymmetric Gaussian
function for $\calF$, ${3\over2}(a_0-a_2\cos^2\thetaB)$ for
$\cos\thetaB$, and an exponential convolved with a Gaussian resolution
function for $\Deltaz$; continuum background distributions are modeled
with an asymmetric Gaussian function for $\calF$,
$(b_0-b_2\cos^2\thetaB)$ for $\cos\thetaB$, and a sum of three Gaussian
functions with a common mean for $\Deltaz$; the coefficients $a_0$,
$a_2$, $b_0$ are close to unity while the coefficient $b_2$ is close to
zero.  

Since $\calF$, $\cosB$ and $\Deltaz$ are independent quantities, we form
a likelihood ratio $\calR=\calLs/(\calLs + \calLc)$ to combine them,
where $\calLs$ and $\calLc$ are products of $\calF$, $\cosB$ (and
$\Deltaz$ if available) likelihood distributions for signal and
continuum, respectively.
The likelihood distribution for the background $\Delta z$ distribution
is determined from data in the sideband region
$5.20\GeVcc<\Mbc<5.24\GeVcc$, $|\DeltaE|<0.3\GeV$; all the other
likelihood distributions are determined from MC samples.

We determine the $\Delta z$ likelihood function separately for the two
datasets as the vertex resolution is improved in the SVD2 with respect
to the SVD1.  As a consequence, we introduce two sets of likelihood
ratios and selection criteria for each decay mode.

On the plane defined by the tagging quality and likelihood ratio, $(r,
\calR)$, signal tends to populate the edges at $r=1$ and $\calR=1$;
continuum tends to populate the edges at $r=0$ and $\calR=0$.  We select
the events in a signal enriched region defined by $\calR > \calR_1$ for
$r>r_1$, and $\calR > 1 - \alpha(1 + r)$ for $r_2<r<r_1$, where the
parameters $r_1$, $r_2$, $R_1$ and $\alpha$ are mode dependent and are
determined so that $N_S/\sqrt{N_S+N_B}$ is maximized (we use this
quantity instead of $N_S/\sqrt{N_B}$ because of the limited statistics
of the MC simulation sample that is used in this procedure).  The values
are $r_1=0.85$, $r_2=0.01$, $R_1>0.8$ and $\alpha=0.025$ for the
$\rhoZG$ mode in the first subset of data; similar values are used for
the other subset and for the $\rhoMG$ and $\omegaG$ modes.  We define
the rest of the area as the background enriched region.

We consider the following $B$ decay backgrounds: $\BtoKG$, other $B\to
X_s\gamma$ processes, $B\to\rho\piZ$ and $\omega\piZ$, $B\to\rho\eta$
and $\omega\eta$, $B^-\to\rho^-\rho^0$, other charmless $B$ decays, and
$b\to c$ backgrounds.  We find the $b\to c$ background to be negligible.
The $\BtoKG$ background may mimic the signal decay $\BtoRG$ if the kaon
from $K^*$ is misidentified as a pion.  In order to further suppress
$\BtoKG$, we calculate $\MKpi$, where the kaon mass is assigned to one
of the pion candidates, and reject the candidate if $\MKpi<0.96$
($0.92$) $\GeVcc$ for the $\rhoZG$ ($\rhoMG$) mode.  The decay chain
$\BtoKBG$, $\Kbar^{*0}\to\KS\piZ$, $\KS\to\piP\piM$ has the same final
state as $\BtoOG$, and has a small contribution due to the tail of the
$K^*$ Breit-Wigner line shape.  In addition, $\BtoKG$ and other
$\BtoXsgamma$ decays contribute to the background when the $\rho$ and
$\omega$ candidates are selected from a random combination of particles.
Charmless decays with a $\piZ$ or $\eta\to\gamma\gamma$, $B\to\rho\piZ$,
$\omega\piZ$, $\rho\eta$ and $\omega\eta$, may mimic the signal if one
of the photons from $\piZ$ or $\eta$ decay is soft and undetected by the
$\piZeta$ veto condition.  To suppress this background, we calculate the
cosine of the helicity angle $\thetahel$, and reject the candidate if
$|\coshel|>0.8$ ($0.6$) for the $\rhoZG$ and $\omegaG$ ($\rhoMG$) modes.
Here, $\thetahel$ is defined as the angle between the $\piP$ and $B$
momentum vectors in the $\rho$ rest frame, or the angle between the
normal to the $\omega$ decay plane and the $B$ momentum vector in the
$\omega$ rest frame.  The decay $B^-\to\rho^-\rho^0$,
$\rho^-\to\pi^-\pi^0$ also contributes to the $\BtoRZG$ mode when both
one pion from the $\rho^-$ decay and one photon from the $\piZ$ decay
are soft and undetected.  The other charmless decays have small
contributions and are considered as an additional background component
when we extract the signal yield.


The reconstruction efficiency for each mode is defined by the fraction
of the signal yield remaining after all selection criteria, where the
signal yield is determined from an extended unbinned maximum likelihood
fit to the MC sample.
The signal distributions are modeled as the product of a Gaussian
function for $\Mbc$ and an empirical function known as the Crystal Ball
line shape \cite{bib:cbls} to reproduce the asymmetric ECL energy
response for $\DeltaE$.  The background component is modeled as the
product of a linear function for $\DeltaE$, and an ARGUS function
\cite{bib:argus-function} for $\Mbc$.  From the fit, we find
efficiencies of about 5\% as listed in Table~\ref{tbl:results}.
The systematic error on the efficiency is the quadratic sum of the
following contributions estimated from control samples: the photon
detection efficiency (2.2\%), measured from radiative Bhabha events;
charged tracking efficiency (1.0\% per track) from partially
reconstructed $D^{*+}\to D^0\piP$, $D^0\to\KS\piP\piM$,
$\KS\to\piP(\piM)$; charged pion identification (1.0\% per pion) from
$D^{*+}\to D^0\piP$, $D^0\to\KM\piP$; neutral pion detection
(4.6--7.3\%) from $\eta$ decays to $\gamma\gamma$, $\piP\piM\piZ$ and
$3\piZ$; $\calR$-$r$ and $\piZeta$ veto requirements (5.4\%) from
$B^-\to D^0\pi^-$, $D^0\to\KM\piP$; and MC statistics (0.9--1.5\%).


We perform an unbinned maximum likelihood fit to the data in the
($\Mbc$, $\DeltaE$) fit region bounded by $|\DeltaE|<0.3\GeV$ and
$\Mbc>5.2\GeVcc$, simultaneously for the three signal modes
(collectively referred to as $\BtoROG$) and the two $\BtoKG$ modes.  We
fit the two data subsets simultaneously, so that in total ten
distributions are included in the fit.  We define the combined branching
fraction $\Br(\BtoROG)=\Br(\BtoRMG)$, assuming the isospin relation
\cite{bib:ali-1994} $\Br(\BtoRMG) = 2\tauBratio\Br(\BtoRBG) =
2\tauBratio\Br(\BtoOG)$, where we use $\tauBratio = 1.086\pm0.017$
\cite{bib:pdg2004}.  We also assume $\Br(\BtoKG)\equiv\Br(\BtoKMG) =
\tauBratio\Br(\BtoKBG)$.

We describe the events in the fit region using a sum of the signal,
continuum, $\KstarG$, and other background hypotheses.  Parameters of
the signal description for $\Mbc$ are calibrated using $\BtoDZpi$
samples.  Those for $\DeltaE$ are calibrated using the result of a fit
to the $\DeltaE$ distribution of the $\BtoKG$ control sample, in which
the mean and the width are allowed to float.  We use the branching
fractions for $\BtoROG$ and $\BtoKG$ as the free parameters, from which
the signal yield for each channel is deduced using the efficiency for
each mode.  We assume the efficiency systematic errors are fully
correlated when we evaluate the systematic error.  The continuum
background is modeled as the product of a linear function for $\DeltaE$
whose slope is allowed to float, and an ARGUS function for $\Mbc$ whose
parameters are fixed from a comparison between data and MC in the
background enriched region and MC in the signal enriched region.  The
continuum contribution in the data sample is allowed to float.  The size
of the $\KstarG$ background component in each $\ROG$ channel is
constrained using the fit to the $\KstarG$ events.  The contributions of
the other backgrounds are fixed using known branching fractions or upper
limits.  The free parameters in the fit are therefore the branching
fractions for $\BtoROG$ and $\BtoKG$, five continuum fractions, and five
continuum $\DeltaE$ slopes.

We also perform individual fits to the three signal modes and the two
$\BtoKG$ modes.  The two data subsets are fitted simultaneously for each
mode.  The size of the $\Kbar^{*0}\gamma$ background in the
$\rho^0\gamma$ mode and $K^{*-}\gamma$ background in the $\rho^-\gamma$
mode are fixed according to the fit results to the $\BtoKG$ modes and
the known particle misidentification probabilities.  The free parameters
in each fit are the signal yield, the continuum fraction, and the
continuum $\DeltaE$ slope.

\begin{figure}[ht]
\begin{center}
\myeps[\figtwoscale]{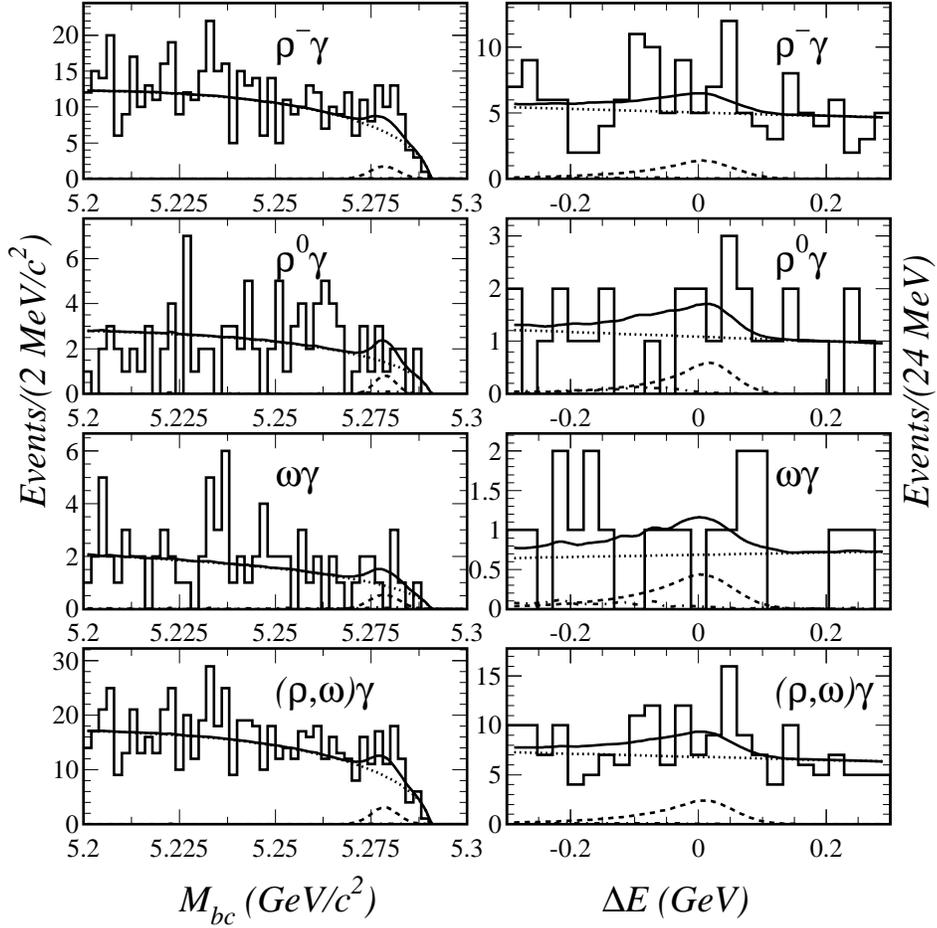}
\caption{Projections of the simultaneous fit results to $\Mbc$ (in the
         region $-0.10\GeV<\DeltaE<0.08\GeV$) and $\DeltaE$ (in the
         region $5.273\GeVcc<\Mbc<5.285\GeVcc$) for the individual modes
         and the sum of them.  Lines represent the total fit result
         (solid), signal (dashed), continuum (dotted), and $B$ decay
         background (dot-dashed) components.}
\label{fig:simfit}
\end{center}
\end{figure}

Results of the simultaneous fit are shown in Fig.~\ref{fig:simfit} and
given in Table~\ref{tbl:results}.  The simultaneous fit gives a
significance of $\sROG$ standard deviations, where the significance is
calculated as $\sqrt{-2\ln(\Lzero/\Lmax)}$, and $\Lmax$ ($\Lzero$) is
the maximum likelihood from the fit when the signal branching fraction
is floated (constrained to be zero).  In order to include the effect of
possible systematic error in the significance calculation, we change
each parameter by one standard deviation in the direction that gives the
smallest resulting significance.  The systematic error in the signal
yield is estimated by varying each of the fixed parameters by its
standard deviation, and then taking the quadratic sum of the deviations
in the signal yield from the nominal value.  The combined branching
fraction is $\Br(\BtoROG)=\BrROG$, where the first and second errors are
statistical and systematic, respectively.

Since the significance is small, we quote a 90\% confidence level upper
limit $\Br_{90}$ using the relation $\int_0^{\Br_{90}}
\calL(x)dx=0.9\int_0^\infty \calL(x)dx$, where $\calL(x)$ is the
likelihood function with the branching fraction fixed at $x$.  The
systematic error is taken into account assuming a Gaussian distribution.
We find
\begin{equation}
\Br(\BtoROG)<\ULROG
\end{equation}
at the 90\% confidence level.  Results of the fits to the
individual modes are also given in Table~\ref{tbl:results}.

A similar fit procedure is performed by using the ratio of branching
fractions $\Br(\BtoROG)/\Br(\BtoKG)$ instead of $\Br(\BtoROG)$, so that
the systematic error partially cancels.  We find
${\Br(\BtoROG)/\Br(\BtoKG)}<\ULROGoverKG$ at the 90\% confidence level.
One can use this result to constrain $\Vtd$: for example, using the
prescription given in Ref.~\cite{bib:ali-2004},
\begin{equation}
 {\Br(\BtoROG)\over\Br(\BtoKG)}=
 \left| {\Vtd\over\Vts} \right|^2
 {(1-m_{(\rho,\omega)}^2/m_B^2)^3 \over (1-m_{K^*}^2/m_B^2)^3}
 \zeta^2
 [1 + \Delta R]
\end{equation}
where the form factor ratio $\zeta=0.85\pm0.10$ and $SU(3)$ breaking
effect $\Delta R=0.1\pm0.1$, we obtain $|\Vtd/\Vts|<\ULVtdoVts$ at the
90\% confidence level.  This limit is consistent with other
determinations of $|\Vtd/\Vts|$~\cite{bib:pdg2004}.


In conclusion, we search for the $\btodgamma$ process using a
simultaneous fit to the $\BtoRG$, $\BtoOG$ and $\BtoKG$ modes.  The
upper limit we obtain is already within the range of SM
predictions~\cite{bib:ali-parkhomenko,bib:bosch-buchalla} and can be
used to constrain $\Vtd$.


We wish to thank the KEKB accelerator group for the excellent
operation of the KEKB accelerator.
We acknowledge support from the Ministry of Education,
Culture, Sports, Science, and Technology of Japan
and the Japan Society for the Promotion of Science;
the Australian Research Council
and the Australian Department of Education, Science and Training;
the National Science Foundation of China under contract No.~10175071;
the Department of Science and Technology of India;
the BK21 program of the Ministry of Education of Korea
and the CHEP SRC program of the Korea Science and Engineering
Foundation;
the Polish State Committee for Scientific Research
under contract No.~2P03B 01324;
the Ministry of Science and Technology of the Russian Federation;
the Ministry of Education, Science and Sport of the Republic of
Slovenia;
the National Science Council and the Ministry of Education of Taiwan;
and the U.S.\ Department of Energy.



\begin{table}[H]
\caption{ Efficiencies, significances, and 90\%
  confidence level upper limits for the branching fractions.}
\label{tbl:results}
\begin{ruledtabular}
\begin{tabular}{lccc}
Mode & Efficiency & Significance 
     & Upper limit \\
{}   & ($\pm$syst) & & (90\% C.L.) \\
\hline
$\BtoROG$ combined &---       &$\sROG$ &$\ULROG$ \\
$\BtoRMG$ &$\effRPG$ &$\sRPG$ &$\ULRPG$ \\
$\BtoRBG$ &$\effRZG$ &$\sRZG$ &$\ULRZG$ \\
$\BtoOG$  &$\effOMG$ &$\sOMG$ &$\ULOMG$ \\
\end{tabular}
\end{ruledtabular}
\end{table}


\begin{thebibliography}{99}

\bibitem{bib:belle-rhogam}
M. Nakao (for the Belle Collaboration), hep-ex/0307031.

\bibitem{bib:babar-rhogam}
BaBar Collaboration, B.~Aubert \etal, \Journal{\PRL}{92}{111801}{2004}.

\bibitem{bib:ali-parkhomenko}
A.~Ali and A.~Ya.~Parkhomenko, \Journal{\EPJC}{23}{89}{2002}.

\bibitem{bib:bosch-buchalla}
In addition to Ref.~\cite{bib:ali-parkhomenko}, for example,
S.~Bosch and G.~Buchalla, \Journal{\NPB}{621}{459}{2002};
T.~Huang, Z.~Li and H.~Zhang, \Journal{\JPG}{25}{1179}{1999}.

\bibitem{bib:kekb}
S.~Kurokawa and E.~Kikutani, \Journal{\NIMA}{499}{1}{2003}.

\bibitem{bib:belle-detector}
Belle Collaboration, A.~Abashian \etal, \Journal{\NIMA}{479}{117}{2002}.

\bibitem{bib:svd2}
Y. Ushiroda, \Journal{\NIMA}{511}{6}{2003}.

\bibitem{bib:belle-pi0pi0} 
Belle Collaboration, S.H.~Lee \etal, \Journal{\PRL}{91}{261801}{2003}.

\bibitem{bib:fisher}
R. A. Fisher, \Journal{Ann. Eugen.}{7}{179}{1936}.

\bibitem{bib:fox-wolfram}
G. C. Fox and S. Wolfram, \Journal{\PRL}{41}{1581}{1978}.

\bibitem{bib:hamlet}
H.~Kakuno \etal, hep-ex/0403022.

\bibitem{bib:cbls}
J.~E.~Gaiser \etal, \Journal{\PRD}{34}{711}{1986}.

\bibitem{bib:argus-function}
ARGUS Collaboration, H.~Albrecht \etal, \Journal{\PLB}{241}{278}{1990}.

\bibitem{bib:ali-1994} 
A. Ali and V. M.~Braun and H.~Simma, \Journal{\ZPC}{6}{437}{1994}.

\bibitem{bib:ali-2004}
A.~Ali, E.~Lunghi, A.~Parkhomenko, \Journal{\PLB}{595}{323}{2004}.

\bibitem{bib:pdg2004}
Particle Data Group, S. Eidelman \etal, \Journal{\PLB}{592}{1}{2004}.

\end{thebibliography}
\end{document}